\documentclass[11pt]{article}
\usepackage{my_aaspp4}
\usepackage{amsmath}
\usepackage{graphics}
\DeclareRobustCommand{\ion}[2]{%
\relax\ifmmode
\ifx\testbx\f@series
{\mathbf{#1\,\mathsc{#2}}}\else
{\mathrm{#1\,\mathsc{#2}}}\fi
\else\textup{#1\,{\mdseries\textsc{#2}}}%
\fi}
\def\kms{km\,s$^{-1}$}
\def\Am{\AA\,mm$^{-1}$}
\def\farcm{\hbox{$.\mkern-4mu^\prime$}}
\def\farcs{\hbox{$.\!\!^{\prime\prime}$}}
\def\degr{\hbox{$^\circ$}}
\def\arcmin{\hbox{$^\prime$}}
\def\arcsec{\hbox{$^{\prime\prime}$}}
\def\hii{\ion{H}{ii}}
\def\feii{\ion{Fe}{ii}}
\def\mgii{\ion{Mg}{ii}}
\begin{document}

\title{ON THE NATURE OF THE FBS BLUE STELLAR OBJECTS AND THE 
COMPLETENESS OF THE BRIGHT QUASAR SURVEY\,\footnotemark}

\footnotetext{\,Partly based on observations collected at the 
Observatoire de Haute-Provence (CNRS, France). The APS databases are supported 
by the National Science Foundation, the National Aeronautics and Space 
Administration, and the University of Minnesota, and are available at 
http://aps.umn.edu/. The Digitized Sky Survey was produced at the Space 
Telescope Science Institute (STScI) under U.S. Government grant NAG W-2166.}

\author{A.M. Mickaelian}
\affil{Byurakan Astrophysical Observatory, Byurakan 378433, Republic of Armenia}
\author{A.C. Gon\c{c}alves, M.P. V\'eron-Cetty, P. V\'eron}
\affil{Observatoire de Haute-Provence (CNRS), 04870 St. Michel l'Observatoire, 
France}

\begin{abstract}
The second part of the First Byurakan Survey is aimed at detecting all bright
($B <$ 16.5) UV-excess starlike objects in a large area of the sky. 
By comparison with other major surveys such as the ROSAT All Sky Survey, 
the ROSAT WGACAT catalogue of point sources, the IRAS survey, the 
6\,cm Green Bank, the 1.4\,GHz NRAO VLA and the 92\,cm Westerbork Northern 
sky surveys and with the catalogue of mean $UBV$ data on stars, we estimate 
the number of AGNs present in the FBS survey and its completeness. 

We have made spectroscopic observations of nine of the most promising FBS
candidates. We have found six new QSOs bringing the total number of known QSOs
in this survey to 42. 

By comparison with the Bright Quasar Survey, we 
found that the completeness of this last survey is of the order of 
70\% rather than 30--50\% as suggested by several authors.

\end{abstract}



\section{INTRODUCTION}

The surface density of bright QSOs ($B <$ 17.0) is still very poorly known. The 
Palomar Green (PG) or Bright Quasar Survey (BQS) [23,53] covering an area of 
10\,714 deg$^{2}$ lead to the discovery of 69 QSOs brighter than 
$M_{B}$ = $-$24 ($H_{\rm o}$ = 50 \kms\,Mpc$^{-1}$) and $B$ = 16.16 
corresponding to 0.0064 deg$^{-2}$. However, several authors [22,33,36,51,58] 
suggested that this survey could be incomplete by a factor 2 to 3.

The First Byurakan Survey (FBS), also known as the Markarian survey, was
carried out in 1965--80 by Markarian et al. [40]. 
It is a slitless spectroscopic photographic survey carried out with the 
40\arcsec\ Schmidt telescope of the Byurakan Observatory. The 1.5\degr\ 
prism used gave a reciprocal dispersion of 1\,800 \Am\ at H$\gamma$. Each 
field is 4\degr$\times$4\degr\ in size. The survey is about 
17\,000 deg$^{2}$ and is complete to about $B$ = 16.5 mag. It has been 
used by Markarian and his collaborators to search for UV excess galaxies; 
1\,500 have been found, including about 10\% Seyfert galaxies and a few 
QSOs. It can also be used for finding UV excess or emission-line star-like 
objects. Such a program -- the second part of the FBS -- has been undertaken 
in 1987 [9,45]. Its main purpose is to take advantage of the large area 
covered to get a reliable estimate of the surface density for bright QSOs. 
The discovery of a number of planetary nebula nuclei, white dwarfs, 
cataclysmic variables (CVs) and other UV excess objects is also expected.

\section{THE FBS SURVEY}

At the present time, 4\,109 deg$^{2}$ have been searched 
(33\degr\ $< \delta <$ 45\degr\ and $\delta >$ 61\degr, excluding 
the Galactic plane) and a catalogue of 1\,103 blue stellar objects has 
been built. It has been published in a series of eleven papers [2-12]. 
It contains 388 objects at $\vert b \vert <$ 30\degr, including 33 at 
12\degr\ $< \vert b \vert <$ 15\degr. Fig.~1 shows the distribution on 
the sky of 
the 1\,103 objects. 433 spectroscopic identifications (397 stars and 36 QSOs) 
are already known, taken mainly from [1,19,35,46,47,55,57,59-65], 
catalogues of spectroscopically identified white dwarfs [41,43], the 
catalogue of cataclysmic variables [17] and the catalogue of subdwarfs [31].

\vspace*{3mm}
\begin{figure}[h]
\begin{center}
\resizebox{16.5cm}{!}{\includegraphics{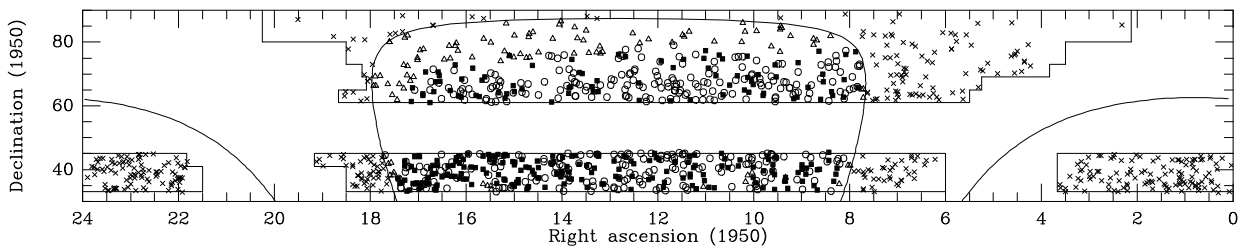}}
\caption{\small Sky map of the FBS objects. Lines of constant galactic 
latitudes are indicated ($b$ = 0\degr\ and $b$ = 30\degr). The nominal 
limits of the FBS are shown. Crosses are for objects at 
$\vert b \vert <$ 30\degr, triangles for objects at $b >$ 30\degr\ but 
outside the PG area, filled squares for objects detected by the PG survey 
and open circles for objects within the PG area, but undetected.}
\end{center}
\end{figure}

106 FBS stars have published $UBV$ colours [23,44] as well as seven 
FBS QSOs [55]. With the exception of FBS 1002+390 ($U-B$ = $-$0.14 [34]), 
they all 
have $U-B$ $<$ $-$0.50 (Fig.~2). The FBS objects have been classified 
as B or N according to the ratio of the intensity of the red and blue 
regions of the spectra. The catalogue contains 862 B objects, 233 N and 
8 others. According to the classification, in general, B objects should 
have a negative $B-V$, while N 
objects should have a positive $B-V$ [9]. In fact, if most N objects 
have a positive $B - V$, a significant fraction of the B objects also have 
a positive $B-V$ (Fig.~2). 

\begin{figure}[h]
\begin{center}
\resizebox{8cm}{!}{\includegraphics{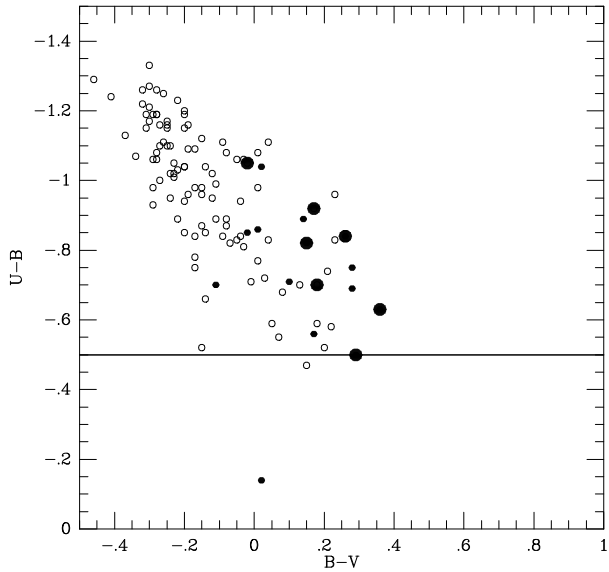}}
\caption{\small Plot of $U-B$ vs. $B-V$ for the FBS objects with 
published photoelectric measurements. Open circles are for B type objects, 
small dots for N types and large dots for QSOs.}
\end{center}
\end{figure}
 
Sixteen QSOs are of B type or 1.8\% of all B 
type objects, while 20 are of N type or 8.5\% of all N type; N type objects 
are therefore much more likely to be QSOs which however have about the 
same probability to be of N or B type. Nevertheless, among the eight FBS 
QSOs with $z >$ 1, seven are of N type, while only one is of B type.

The positional accuracy for the first 429 objects (first four papers) is quoted
as being about 1\arcmin. Later the accuracy was increased to 0\farcm5 [2].
An accurate optical position has been measured for 195 objects on the Digitized
Sky Survey; the accuracy is $\sim$0\farcs6 [54]. These positions are given 
in Table 1. We have compared the published FBS positions with sub-arcsecond 
accuracy positions for 104 objects of the first four papers and 117 objects 
of the last seven papers. After correcting for printing errors in the 
published positions of FBS 0649+716, 0935+416, 1559+369 and 
1619+648, we found 
$\sigma_{\alpha}$ = 46\arcsec, $\sigma_{\delta}$ = 40\arcsec\ and 
$\sigma_{\alpha}$ = 9\arcsec, $\sigma_{\delta}$ = 7\arcsec\ respectively; 
these errors are substantially smaller than the initial estimates.

The magnitudes of the objects in the first four papers have been measured from 
the $O$ Palomar Sky Survey prints, using the relationship between stellar 
diameters and magnitudes established by [26]; however in these papers, the PG 
magnitude, when available, was given rather than the FBS magnitude. 
In contrast, 
for the objects listed in the last seven papers, the magnitudes have been 
measured on both the $O$ and $E$ prints, with the calibration given by [32] and 
the $V$ magnitude computed with the formula :  $V = E-(O-E)/$3.2.  
The photometric accuracy, which was thought to be 0.5 mag in the first series 
of measurements, was improved to 0.3 mag starting with the fifth paper [2].

\begin{table}
\begin{center}
\caption{\small Accurate positions for 195 FBS objects (continues).} 
\vspace{0.5cm} 
\begin{tabular}{p{3.5cm}crcr}
\tableline 
\tableline
Name             & & $\alpha$ (B1950) & & $\delta$ (B1950) \\
\tableline    
 FBS 0004+330    & & 00 04 57.58   & & 33 00 49.2 \\ 	
 FBS 0019+348    & & 00 19 10.52   & & 34 48 13.3 \\
 FBS 0019+401    & & 00 19 44.22   & & 40 09 12.4 \\
 FBS 0028+441    & & 00 28 01.72   & & 44 07 59.3 \\
 FBS 0028+435    & & 00 28 53.76   & & 43 32 32.1 \\ 
 FBS 0038+431    & & 00 38 10.80   & & 43 08 32.7 \\
 FBS 0043+343    & & 00 43 09.54   & & 34 18 11.0 \\
 FBS 0047+347    & & 00 47 14.22   & & 34 41 52.7 \\
 FBS 0051+417    & & 00 51 36.77   & & 41 47 08.8 \\
 FBS 0058+431    & & 00 58 18.96   & & 43 07 17.9 \\	
 FBS 0125+351    & & 01 25 14.97   & & 35 06 10.5 \\
 FBS 0125+386    & & 01 25 35.36   & & 38 39 07.4 \\
 FBS 0127+408    & & 01 27 00.52   & & 40 47 19.5 \\
 FBS 0140+427    & & 01 39 59.78   & & 42 42 14.6 \\
 FBS 0150+396    & & 01 50 06.88   & & 39 41 00.0 \\
 FBS 0154+391    & & 01 54 18.51   & & 39 08 25.2 \\
 FBS 0156+439    & & 01 56 40.53   & & 43 58 47.4 \\
 FBS 0212+385    & & 02 12 55.08   & & 38 32 26.6 \\
 FBS 0217+343    & & 02 17 17.08   & & 34 20 00.4 \\
 FBS 0228+447    & & 02 28 42.59   & & 44 44 13.1 \\	
 FBS 0233+373    & & 02 33 37.67   & & 37 21 17.7 \\
 FBS 0255+379    & & 02 55 30.67   & & 37 57 40.1 \\
 FBS 0306+333    & & 03 06 07.77   & & 33 20 01.4 \\
 FBS 0315+417    & & 03 15 04.13   & & 41 44 27.0 \\
 FBS 0421+740    & & 04 21 26.14   & & 74 00 46.2 \\
 FBS 0432+763    & & 04 32 27.59   & & 76 18 44.6 \\
 FBS 0437+756    & & 04 37 51.88   & & 75 33 21.5 \\
 FBS 0613+431    & & 06 13 06.93   & & 43 10 59.2 \\
 FBS 0614+769    & & 06 14 01.33   & & 76 52 53.3 \\
 FBS 0624+428    & & 06 24 17.31   & & 42 48 39.1 \\	
 FBS 0632+663    & & 06 32 03.44   & & 66 15 21.6 \\
 FBS 0637+786    & & 06 37 33.69   & & 78 38 04.1 \\
 FBS 0639+391    & & 06 39 14.98   & & 39 11 22.0 \\	
\tableline
\tableline
\end{tabular}
\end{center}
\end{table}
\addtocounter{table}{-1}
\begin{table}
\begin{center}
\caption{\small Accurate positions for 195 FBS objects (continued).} 
\vspace{0.5cm} 
\begin{tabular}{p{3.5cm}crcr}
\tableline 
\tableline
Name             & & $\alpha$ (B1950) & & $\delta$ (B1950) \\
\tableline    
 FBS 0649+716    & & 06 49 08.05   & & 71 37 22.5 \\	
 FBS 0652+799    & & 06 52 50.66   & & 79 55 54.5 \\
 FBS 0654+366    & & 06 54 40.40   & & 36 34 23.4 \\
 FBS 0702+616    & & 07 02 14.87   & & 61 38 29.6 \\
 FBS 0706+407    & & 07 06 42.65   & & 40 41 18.2 \\
 FBS 0716+365    & & 07 16 52.62   & & 36 29 01.9 \\
 FBS 0732+396    & & 07 32 58.96   & & 39 32 59.4 \\
 FBS 0742+653    & & 07 42 41.23   & & 65 20 24.7 \\
 FBS 0742+337    & & 07 42 59.53   & & 33 40 29.4 \\
 FBS 0744+818    & & 07 44 27.07   & & 81 49 30.3 \\
 FBS 0744+652    & & 07 44 52.30   & & 65 10 16.2 \\
 FBS 0747+729    & & 07 47 53.47   & & 72 57 44.2 \\
 FBS 0749+725    & & 07 49 44.82   & & 72 32 17.9 \\
 FBS 0752+769    & & 07 52 40.64   & & 76 54 28.6 \\
 FBS 0808+435    & & 08 08 05.06   & & 43 32 13.8 \\
 FBS 0808+628    & & 08 08 05.58   & & 62 45 24.2 \\
 FBS 0819+364    & & 08 19 32.57   & & 36 23 54.1 \\
 FBS 0821+676    & & 08 21 33.89   & & 67 37 07.7 \\
 FBS 0827+738    & & 08 27 10.40   & & 73 47 17.6 \\
 FBS 0836+619    & & 08 36 33.96   & & 61 58 32.2 \\
 FBS 0845+812    & & 08 45 25.59   & & 81 10 16.7 \\
 FBS 0848+437    & & 08 48 22.96   & & 43 45 22.7 \\
 FBS 0850+639    & & 08 50 37.98   & & 63 54 46.1 \\
 FBS 0904+643    & & 09 04 58.96   & & 64 22 06.4 \\
 FBS 0906+368    & & 09 06 18.10   & & 36 49 46.8 \\
 FBS 0914+656    & & 09 14 51.16   & & 65 36 45.7 \\
 FBS 0920+674    & & 09 20 01.27   & & 67 23 05.9 \\
 FBS 0920+366    & & 09 20 49.65   & & 36 36 35.2 \\
 FBS 0924+732    & & 09 24 23.00   & & 73 09 46.9 \\
 FBS 0926+850    & & 09 26 34.50   & & 85 01 39.0 \\
 FBS 0929+733    & & 09 29 12.25   & & 73 16 16.2 \\
 FBS 0932+437    & & 09 32 12.72   & & 43 44 31.0 \\	
\tableline
\tableline
\end{tabular}
\end{center}
\end{table}
\addtocounter{table}{-1}
\begin{table}
\begin{center}
\caption{\small Accurate positions for 195 FBS objects (continued).} 
\vspace{0.5cm} 
\begin{tabular}{p{3.5cm}crcr}
\tableline 
\tableline
Name             & & $\alpha$ (B1950) & & $\delta$ (B1950) \\
\tableline    
 FBS 0933+614    & & 09 33 53.15   & & 61 25 18.0 \\	
 FBS 0935+679    & & 09 35 18.71   & & 67 54 00.8 \\
 FBS 0935+395    & & 09 35 42.66   & & 39 32 25.3 \\
 FBS 0938+374    & & 09 38 24.45   & & 37 26 03.6 \\
 FBS 0938+447    & & 09 38 34.70   & & 44 42 46.2 \\
 FBS 0941+664    & & 09 41 18.63   & & 66 25 20.8 \\
 FBS 0944+713    & & 09 44 45.41   & & 71 15 09.4 \\
 FBS 0950+664    & & 09 50 09.49   & & 66 22 30.9 \\
 FBS 0953+686    & & 09 53 03.57   & & 68 36 35.8 \\
 FBS 0954+697    & & 09 54 24.61   & & 69 43 20.5 \\
 FBS 0958+353    & & 09 58 17.44   & & 35 19 37.2 \\
 FBS 1002+437    & & 10 02 37.27   & & 43 47 17.2 \\
 FBS 1003+678    & & 10 03 08.82   & & 67 47 25.6 \\
 FBS 1007+382    & & 10 08 04.69   & & 38 16 48.9 \\
 FBS 1040+451    & & 10 40 36.64   & & 45 09 14.0 \\
 FBS 1054+436    & & 10 54 35.66   & & 43 37 10.6 \\
 FBS 1057+719    & & 10 57 07.54   & & 71 54 10.6 \\
 FBS 1102+347    & & 11 02 54.97   & & 34 41 47.0 \\
 FBS 1103+385    & & 11 03 04.38   & & 38 29 16.9 \\
 FBS 1104+408    & & 11 04 54.15   & & 40 49 08.5 \\
 FBS 1108+402    & & 11 08 10.48   & & 40 15 35.6 \\
 FBS 1112+668    & & 11 12 19.28   & & 66 48 23.4 \\
 FBS 1122+426    & & 11 22 09.44   & & 42 41 53.7 \\
 FBS 1125+634    & & 11 25 45.85   & & 63 21 16.2 \\
 FBS 1129+823    & & 11 29 34.93   & & 82 19 40.7 \\
 FBS 1133+754    & & 11 33 32.66   & & 75 23 30.2 \\ 
 FBS 1138+648\,A & & 11 38 45.38   & & 64 49 08.6 \\
 FBS 1138+648\,B & & 11 38 46.11   & & 64 49 12.1 \\ 
 FBS 1139+437    & & 11 39 35.91   & & 43 40 54.9 \\
 FBS 1140+719    & & 11 40 48.57   & & 71 57 58.5 \\
 FBS 1141+406    & & 11 41 40.55   & & 40 41 08.0 \\
 FBS 1147+673    & & 11 47 46.03   & & 67 15 28.5 \\
 FBS 1148+444    & & 11 48 47.18   & & 44 29 23.1 \\	
\tableline
\tableline
\end{tabular}
\end{center}
\end{table}
\addtocounter{table}{-1}
\begin{table}
\begin{center}
\caption{\small Accurate positions for 195 FBS objects (continued).} 
\vspace{0.5cm} 
\begin{tabular}{p{3.5cm}crcr}
\tableline 
\tableline
Name             & & $\alpha$ (B1950) & & $\delta$ (B1950) \\
\tableline    
 FBS 1149+394    & & 11 49 27.95   & & 39 25 08.7 \\	
 FBS 1150+334    & & 11 50 16.53   & & 33 23 59.8 \\
 FBS 1156+432    & & 11 56 13.19   & & 43 15 48.2 \\
 FBS 1201+437    & & 12 01 51.15   & & 43 47 39.3 \\
 FBS 1211+393    & & 12 11 04.32   & & 39 17 34.8 \\
 FBS 1223+665    & & 12 23 13.26   & & 66 31 26.9 \\
 FBS 1229+383    & & 12 29 03.27   & & 38 19 15.0 \\
 FBS 1229+710    & & 12 29 28.28   & & 71 00 47.2 \\
 FBS 1230+417    & & 12 30 00.95   & & 41 45 51.3 \\
 FBS 1231+828    & & 12 31 46.59   & & 82 50 21.8 \\
 FBS 1232+379    & & 12 32 28.31   & & 37 54 14.5 \\
 FBS 1235+699    & & 12 35 12.88   & & 69 58 13.2 \\
 FBS 1240+631    & & 12 40 27.96   & & 63 06 21.5 \\
 FBS 1248+374    & & 12 48 44.76   & & 37 23 00.3 \\
 FBS 1249+433    & & 12 49 48.58   & & 43 20 24.5 \\
 FBS 1255+447    & & 12 55 01.75   & & 44 45 46.8 \\
 FBS 1311+664    & & 13 11 50.12   & & 66 27 01.9 \\ 
 FBS 1315+645    & & 13 15 09.88   & & 64 31 09.8 \\
 FBS 1315+447    & & 13 15 49.63   & & 44 43 19.5 \\
 FBS 1316+446    & & 13 16 01.31   & & 44 40 06.0 \\
 FBS 1324+448    & & 13 24 54.58   & & 44 50 36.4 \\
 FBS 1335+369    & & 13 35 38.77   & & 36 52 50.6 \\
 FBS 1338+666    & & 13 38 03.38   & & 66 35 50.6 \\
 FBS 1340+813    & & 13 40 40.59   & & 81 18 10.8 \\
 FBS 1351+640    & & 13 51 46.29   & & 64 00 29.0 \\
 FBS 1352+386    & & 13 52 26.75   & & 38 39 18.4 \\
 FBS 1352+451    & & 13 52 49.55   & & 45 08 13.9 \\
 FBS 1356+389    & & 13 56 24.73   & & 38 58 27.6 \\
 FBS 1359+411    & & 13 59 12.50   & & 41 09 01.9 \\
 FBS 1401+865    & & 14 01 11.14   & & 86 29 42.8 \\
 FBS 1402+436    & & 14 02 37.67   & & 43 41 26.9 \\
 FBS 1413+757    & & 14 13 08.72   & & 75 40 15.8 \\	
\tableline
\tableline
\end{tabular}
\end{center}
\end{table}
\addtocounter{table}{-1}
\begin{table}
\begin{center}
\caption{\small Accurate positions for 195 FBS objects (continued).} 
\vspace{0.5cm} 
\begin{tabular}{p{3.5cm}crcr}
\tableline 
\tableline
Name             & & $\alpha$ (B1950) & & $\delta$ (B1950) \\
\tableline    
 FBS 1429+373    & & 14 29 54.34   & & 37 19 41.7 \\	
 FBS 1437+398    & & 14 37 18.97   & & 39 49 35.3 \\
 FBS 1440+753    & & 14 40 14.76   & & 75 18 20.0 \\
 FBS 1444+637    & & 14 44 57.08   & & 63 41 53.2 \\
 FBS 1449+440    & & 14 49 36.48   & & 44 06 03.6 \\
 FBS 1449+642    & & 14 49 37.67   & & 64 15 46.8 \\
 FBS 1452+762    & & 14 52 16.36   & & 76 12 10.1 \\
 FBS 1500+752    & & 15 00 43.03   & & 75 10 33.1 \\
 FBS 1501+664    & & 15 01 24.25   & & 66 24 01.4 \\
 FBS 1513+442    & & 15 13 02.08   & & 44 12 40.7 \\
 FBS 1522+663    & & 15 22 16.88   & & 66 15 31.0 \\
 FBS 1523+363    & & 15 23 16.94   & & 36 15 38.0 \\
 FBS 1534+389    & & 15 34 32.03   & & 38 55 52.6 \\
 FBS 1539+355    & & 15 36 02.86   & & 35 28 08.8 \\
 FBS 1551+719    & & 15 51 40.69   & & 71 54 05.0 \\
 FBS 1554+403    & & 15 54 04.46   & & 40 20 24.6 \\
 FBS 1557+448    & & 15 57 08.98   & & 44 49 30.8 \\
 FBS 1559+369    & & 15 59 32.42   & & 36 57 20.2 \\
 FBS 1602+408    & & 16 02 43.08   & & 40 49 06.3 \\
 FBS 1603+369    & & 16 03 43.62   & & 36 57 42.3 \\
 FBS 1605+684    & & 16 05 29.76   & & 68 22 07.7 \\
 FBS 1605+627    & & 16 05 47.96   & & 62 40 55.8 \\
 FBS 1607+439    & & 16 07 53.88   & & 43 54 10.5 \\
 FBS 1619+749    & & 16 19 47.45   & & 74 55 38.0 \\
 FBS 1619+648    & & 16 19 55.70   & & 64 43 01.2 \\
 FBS 1634+706    & & 16 34 51.56   & & 70 37 37.5 \\
 FBS 1636+351    & & 16 36 36.52   & & 35 06 03.8 \\
 FBS 1638+388    & & 16 38 34.72   & & 38 48 04.0 \\
 FBS 1640+362    & & 16 40 08.90   & & 36 09 43.2 \\
 FBS 1641+399    & & 16 41 17.55   & & 39 54 10.7 \\
 FBS 1641+388    & & 16 41 18.89   & & 38 46 42.4 \\
 FBS 1648+371    & & 16 48 22.63   & & 37 06 16.4 \\
 FBS 1648+407    & & 16 48 40.95   & & 40 42 25.3 \\	
\tableline
\tableline
\end{tabular}
\end{center}
\end{table}
\addtocounter{table}{-1}
\begin{table}
\begin{center}
\caption{\small Accurate positions for 195 FBS objects (end).} 
\vspace{0.5cm} 
\begin{tabular}{p{3.5cm}crcr}
\tableline 
\tableline
Name             & & $\alpha$ (B1950) & & $\delta$ (B1950) \\
\tableline    
 FBS 1656+354    & & 16 56 01.70   & & 35 25 05.1 \\	
 FBS 1657+344    & & 16 57 01.28   & & 34 23 23.1 \\	
 FBS 1658+440    & & 16 58 17.09   & & 44 05 23.6 \\
 FBS 1715+409    & & 17 15 29.57   & & 40 39 44.6 \\
 FBS 1715+424    & & 17 15 45.05   & & 42 29 18.8 \\
 FBS 1716+394    & & 17 16 22.39   & & 39 19 49.3 \\
 FBS 1722+356    & & 17 22 48.91   & & 35 36 55.3 \\
 FBS 1743+440    & & 17 43 26.45   & & 44 05 51.0 \\
 FBS 1745+420    & & 17 44 55.66   & & 42 04 44.0 \\
 FBS 1755+663    & & 17 55 41.53   & & 66 19 16.4 \\
 FBS 1756+394    & & 17 55 55.40   & & 39 21 14.2 \\
 FBS 1756+441    & & 17 56 11.82   & & 44 11 07.8 \\
 FBS 1756+352    & & 17 56 30.04   & & 35 09 17.3 \\
 FBS 1800+686    & & 18 00 26.61   & & 68 35 56.0 \\
 FBS 1810+374    & & 18 10 39.33   & & 37 24 40.4 \\
 FBS 1815+381    & & 18 15 39.96   & & 38 09 43.1 \\
 FBS 1820+809    & & 18 20 54.40   & & 80 54 13.7 \\
 FBS 1821+643    & & 18 21 36.76   & & 64 20 18.7 \\
 FBS 1822+352    & & 18 22 21.32   & & 35 14 38.3 \\
 FBS 1822+414    & & 18 22 21.69   & & 41 27 33.2 \\
 FBS 1833+447    & & 18 33 25.29   & & 44 45 48.6 \\
 FBS 1833+434    & & 18 33 45.70   & & 43 25 00.7 \\
 FBS 2149+425    & & 21 49 04.38   & & 42 32 39.1 \\
 FBS 2152+408    & & 21 52 47.31   & & 40 49 58.0 \\
 FBS 2212+421    & & 22 12 29.04   & & 42 08 08.5 \\
 FBS 2246+414    & & 22 47 00.80   & & 41 28 01.1 \\
 FBS 2248+446    & & 22 48 19.36   & & 44 41 14.5 \\
 FBS 2249+391    & & 22 49 45.92   & & 39 05 20.2 \\
 FBS 2302+427    & & 23 02 43.64   & & 42 46 33.8 \\
 FBS 2308+425    & & 23 08 26.99   & & 42 33 51.2 \\
 FBS 2315+443    & & 23 15 48.38   & & 44 20 01.2 \\
 FBS 2340+422    & & 23 40 54.17   & & 42 17 39.9 \\	
\tableline
\tableline
\end{tabular}
\end{center}
\end{table}

Comparison of the FBS and photoelectric $V$ magnitudes for the objects in the
last seven papers gives $\langle FBS - V \rangle$ = 0.06 mag and 
$\sigma$ = 0.46 mag. For 29 objects from the first four papers, we 
have $\langle FBS - V \rangle$ = $-$0.40 mag, $\sigma$ = 0.75 mag. 
This confirms the significant improvement of the photometric accuracy achieved
in the second part of the catalogue. However, the accuracy seems to be 0.75 mag 
and 0.45 mag in the first and second part respectively, rather than the 
estimated 0.5 mag and 0.3 mag.

\section{COMPARISON WITH X-RAY, INFRARED AND RADIO SURVEYS}

\subsection{The ROSAT All-Sky Survey Bright Source Catalogue (RASS-BSC)}
We have cross-correlated our list of 1\,103 UV-excess objects with the ROSAT 
All-Sky Survey bright source catalogue [56]. There are 2\,225 X-ray sources in 
the area of interest. We have found 57 X-ray sources within 4\arcmin\ of a FBS 
source, while we expected eight chance coincidences. Therefore most of them 
are probably real associations. We have measured the accurate optical 
position of 
the 33 non-QSO coincidences (Table 1); the differences between the X-ray and 
accurate optical positions are smaller than 40\arcsec\ except for eight 
objects, confirming the reality of the associations.

The source RX J17173+4227, which is associated with the radiosource 
B3 1715+425, has been identified with the Zwicky cluster of galaxies 
Zw 8193 at $z$ = 0.183 [13]; the X-ray and radio positions are in good 
agreement within the error limits, while the optical position of 
FBS 1715+424 is about 42\arcsec\ away from the X-ray position; this is 
therefore a chance coincidence. Within 1\arcmin, we found 50 coincidences: 
21 QSOs, 22 stars and seven unidentified objects listed in Table 2; 
recently, one of these objects (FBS 0950+664) has been identified on 
objective prism plates as an AGN [14].

\subsection{The ROSAT WGACAT catalogue of point sources}
The WGACAT catalogue has been generated using the ROSAT PSPC pointed data
publicly available as of September 1994. It contains more than 45\,600 
individual sources in a total of 2\,624 fields [66]. 13\,937 sources are 
located within the FBS area; we have cross-correlated this list with the FBS. 
There are 53 X-ray sources within 2\arcmin\ of a FBS object, while we expect 
13 chance coincidences. 28 of them are also listed in the RASS catalogue. 
We have measured accurate optical positions for the 25 remaining objects 
(Table 1) and recomputed the separation with the companion X-ray sources. 
There are 13 FBS objects within 30\arcsec\ from an X-ray source (excluding 
the sources appearing in the RASS catalogue). Seven are QSOs, four are 
stars (including 3 CVs) and only two (listed in Table 2) were 
of unknown nature before our spectroscopic observations.

\begin{table}[h]
\begin{center}
\caption{\small FBS coincidences with ROSAT and VLA sources.}
\vspace{2.8mm}
\begin{tabular}{p{2cm}cccllcrcl}
\tableline
\tableline
 FBS name &  Mag &  $O$  &   & (1)    & (2)    &   & $b\:\:$ &      & $\:\:z$ \\
\tableline
 0732+396 & 16.0 & 14.70 & X & 10     & 20     & N &  25.1 & QSO  & 0.118   \\
 0950+664 & 16.7 & 17.00 & X & 15     & 16     & Y &  42.4 & AGN  &         \\
 1112+668 & 17.0 & 16.53 & X & 10     & \,~4   & Y &  47.9 & QSO  & 0.544   \\
 1150+334 & 16.2 & 16.30 & R & \,~0.8 & \,~1.2 & Y &  76.0 & QSO  & 1.40    \\
 1235+699 & 17.9 & 17.96 & x & \,~5   & \,~4   & Y &  47.4 & QSO  & 0.522 ? \\
 1255+447 & 16.5 & 16.48 & X & 10     & 13     & Y &  72.6 & QSO  & 0.300   \\
 1315+447 & 17.0 & 17.33 & x & \,~5   & 15     & Y &  71.9 & DZ:  &         \\
 1324+448 & 17.0 & 18.09 & X & \,~8   & \,~9   & Y &  71.1 & QSO  & 0.331 ? \\
 1500+752 & 16.9 &  ---  & X & 10     & 11     & N &  39.5 & DA:  &         \\
 1822+352 & 15.8 &  ---  & R & \,~3.5 & \,~3.9 & N &  20.5 & DA   &         \\
 2308+425 & 13.5 &  ---  & X & 12     & \,~8   & N &$-$16.3&      &         \\
\tableline
\tableline
\end{tabular}
\vspace*{3mm}
\begin{tabular}{p{12.5cm}}
\footnotesize
X: in the ROSAT All Sky Survey Bright Source Catalogue; x: in the ROSAT WGACAT
Catalogue; R: in the NRAO VLA Sky Survey; Y: in the PG area; N: not in the 
PG area; O: APS $O$ magnitudes; (1): error of the ROSAT or VLA
position (in arcsec); (2): distance between the FBS and ROSAT or VLA positions
(in arcsec).
\end{tabular}
\end{center}
\end{table}

\subsection{The IRAS point source catalogue}

We have cross-correlated the IRAS point source catalogue [15] with the FBS.
There are 10\,537 IRAS sources in the area of interest. We expect ten chance
coincidences within 2\arcmin\ from the 1\,103 FBS objects and 2.5 within 
1\arcmin. We found 17 IRAS sources within 2\arcmin\ of the FBS objects. We 
have measured accurate optical positions for these objects (Table 1); 
two turned out to be more than 2\arcmin\ from the IRAS sources; one 
(FBS 1340+813), a white dwarf, is located near the bright K2 star 
SAO 2257 which is identified with the source IRAS 13407+8118 [8]. 
FBS 0432+763 is located near the M star SAO 5262 which is associated with 
the IRAS source. IRAS 04378+7532 is associated with the galaxy UGC 3130 
rather than with FBS 0437+756. IRAS 16402+3611 is associated with the M 
star CLS 106 [52] rather than with FBS 1640+362. IRAS 17562+4412 coincides 
with an uncatalogued bright star and is therefore probably not associated 
with FBS 1756+441. These seven objects are certainly chance coincidences.

All the four FBS planetary nebulae are detected by IRAS, while none of the UV
excess stars is. Five QSOs are also detected, i.e. only $\sim$14\% of 
all known QSOs in the field. The position of FBS 1821+643 coincides with that 
of the nucleus of the planetary nebula PK 094+27.1 [4]; it is not the 
QSO KUV 18217+6419 which is located 84\arcsec\ away. There is only one 
unclassified FBS object coinciding with an IR source, IRAS 11334+7523 
which however is located 14\arcsec\ away from a 16$^{\rm th}$ mag galaxy, 
a more likely identification. 

The small fraction of all QSOs 
which are detected as IRAS sources make this survey of little use 
to check the completeness of the FBS.

\subsection{The Green Bank 6\,cm (GB6) radio survey}

The GB6 survey [24] covers the declination band 0\degr\ $< \delta <$ 75\degr. 
It contains 54\,579 sources stronger than 25\,mJy. 16\,050 sources are within 
the FBS area. We expect 15 chance coincidences within 2\arcmin\ and four 
within 1\arcmin. We found 18 and 12 coincidences within 2\arcmin\ and 1\arcmin, 
respectively. Eight known QSOs are detected, all within 30\arcsec\ of the 
radio-positions. We have measured accurate optical positions (Table 1) for 
the ten remaining objects not known to be QSOs. We are left with three objects 
for which the distance between optical and radio positions is less than three 
times the quoted radio error. Two (FBS 0958+353 and 1534+389) are stars. A 
more accurate position of the third source associated with FBS 1619+749, 
measured with the NRAO VLA sky survey (see below), excludes this 
identification.

\subsection{The NRAO VLA sky survey}

The NRAO VLA Sky Survey (NVSS) covers the sky north of J\,2000 
$\delta$ = $-$40\degr\ (82\% of the celestial sphere) at 1.4 GHz [16]. 
It contains almost 2\,10$^{6}$ discrete sources stronger than $S$ = 2.5\,mJy. 
The rms positional uncertainties vary from $<$ 1\arcsec\ for sources stronger 
than 15\,mJy to 7\arcsec\ at the survey limit. The source surface density is 
about 60 deg$^{-2}$.

We have searched the NVSS catalogue for sources within 2\arcmin\ from the FBS
objects in the first four papers and within 1\arcmin\ in the last seven papers,
excluding the known stars, in all 705 objects; 13 lie in as yet uncatalogued
regions of the NVSS survey. We have found 54 coincidences; we have measured the
accurate optical positions of these FBS objects (Table 1). Using these new
optical positions, and excluding those objects for which the distance between
radio and optical positions exceeds three times the radio-position error, we 
are left with ten coincidences, including eight known QSOs (three known 
extended radio quasars: B2 1512+37, 3C 249.1 and 3C 263.0 have been excluded 
by this procedure). The two new probable radio-identifications 
(listed in Table 2) are FBS 1150+334 and 1822+352.

\subsection{The Westerbork Northern Sky Survey}
The Westerbork Northern Sky Survey (WENSS) is a low-frequency radio-survey that
covers the whole sky north of $\delta$ = 30\degr\ at a wavelength of 92\,cm 
to a
limiting flux density of approximately 18\,mJy [50]. The WENSS comprises two
source catalogues: the main catalogue contains 211\,234 sources in the 
declination range 28\degr\ $< \delta <$ 76\degr\ (83\,134 in the FBS area); 
the polar catalogue contains 18\,186 sources above 72\degr\ (12\,239 in the 
FBS area). The positional accuracy ranges from 1\farcs5 for the brighter 
sources to 10\arcsec\ for the weakest. The source surface density is about 
23 deg$^{-2}$. The total number of coincidences within 1\arcmin\ with FBS 
objects not known to be stars is equal to 19 (including eight known QSOs), 
while 15 are expected by chance. The distances between optical and radio 
positions for the eight QSOs are all smaller than 20\arcsec; for the 
unclassified objects, the smallest distance is 34\arcsec, suggesting that no
new QSO has been detected.
                            
\section{OBSERVATIONS AND DATA REDUCTION}

Spectroscopic observations of nine of the twelve objects associated either
with an X-ray or a radio source (Table 2) and of 22 other FBS objects were
carried out on October 27 and 28, 1997 and on May 25 and 26, 1998 with the
CARELEC spectrograph [39] attached to the Cassegrain focus of the Observatoire
de Haute-Provence (OHP) 1.93\,m telescope. A 260\,\Am\ grating was used; the 
spectral range was 3810--7365\,\AA. The detector was a 512$\times$512 pixels, 
27$\times$27\,$\mu$m Tektronix CCD. The slit width was 2\farcs1, 
corresponding to a projected slit width on the detector of 52\,$\mu$m, 
or 1.9 pixel. The resolution, as measured on the night sky emission lines, 
was 14.3\,\AA\ FWHM. The spectra were flux calibrated using the standard 
stars EG 145 and Feige 66 [42] which were also used to correct the 
observations for the atmospheric absorption.

Six of the X-ray or radio sources turned out to be QSOs, while three are stars. 
The 22 other observed objects are all stars, except one which is a \hii\ 
galaxy. Spectra of the nine X-ray or radio sources observed are presented 
in Fig.~3. 

The journal of observations is given in Table 3, together with relevant data.

\begin{figure}[h]
\begin{center}
\resizebox{16.5cm}{!}{\includegraphics{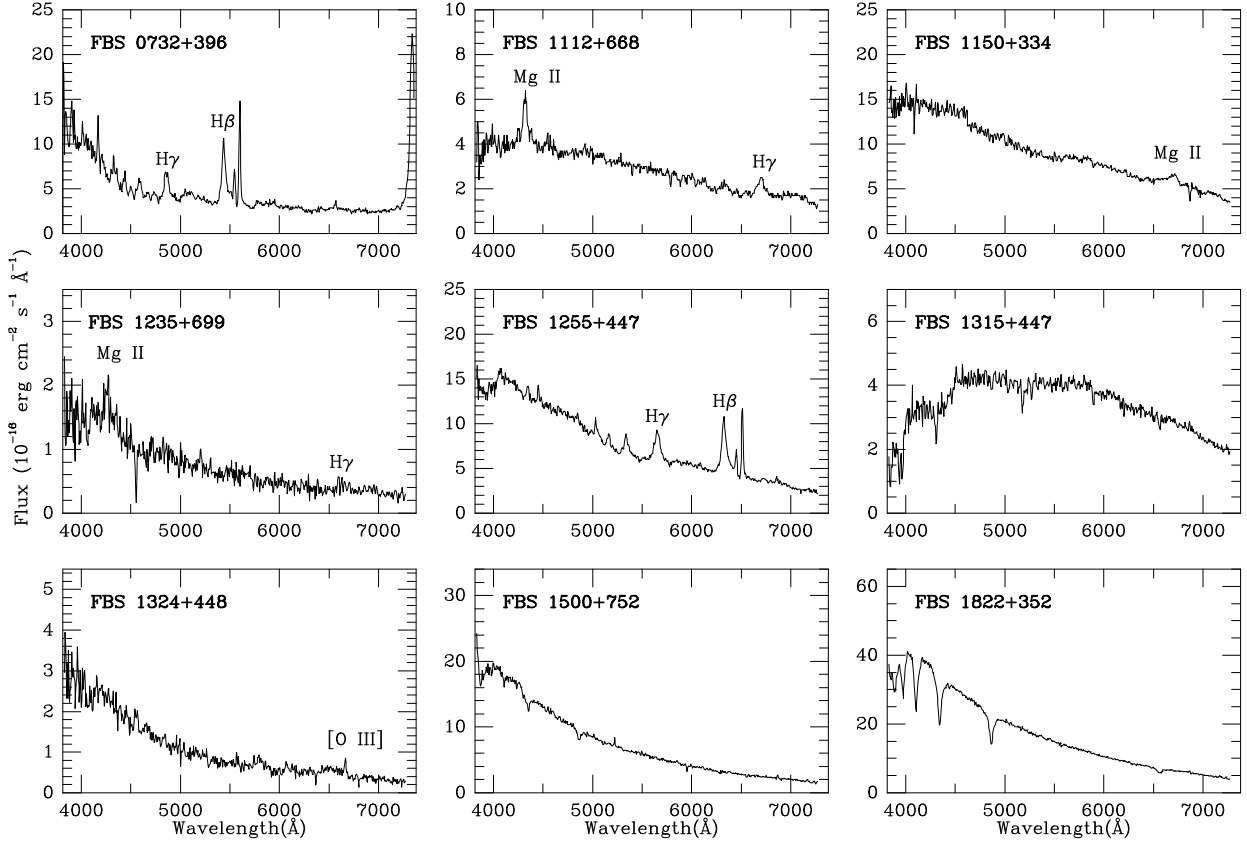}}
\caption{\small Spectra of the nine observed FBS objects identified 
either with an X-ray or radio source.}
\end{center}
\end{figure}

{\bf FBS 0732+396} was suspected of being non-stellar and having emission 
lines [9]. Our spectrum (Fig.~3) shows emission lines characteristic of a 
NLS1 galaxy at $z$ = 0.118, with relatively narrow Balmer lines and strong 
\feii\ lines.

{\bf FBS 1255+447} is HS 1255+4445 at $z$ = 0.30 [25].
 
\section{DISCUSSION}

\subsection{Completeness of the FBS}
The Catalogue of mean $UBV$ data on stars [44] contains 102 stars in the FBS
area, with 11.0 $< V <$ 16.5 (bright stars are saturated 
on the FBS plates and are therefore missed) and $U-B$ $<$ $-$0.50; 
53 are included in the FBS catalogue suggesting that, in this magnitude 
interval, the completeness of the FBS is 52\% (53/102). Only 9\% (2/23) 
of the stars weaker than $V$ = 16.5 appear in this catalogue. The survey 
is insensitive to objects with $U-B$ $>$ $-$0.5 (0/9); 
its completeness increases from $\sim$20\% at $U-B$ $\sim$ $-$0.6 to 
$\sim$80\% for $U-B$ $<$ $-$1.0.

The QSO $U-B$ colour changes with $z$; but most of the changes, at least for 
$z$ $<$ 2.2, are due to the presence of an emission line in one of 
the two filters. The $U-B$ colour of the continuum is in the range 
$-$0.9 $<$ $U-B$ $<$ $-$0.7. Slitless spectroscopic surveys are sensitive 
to the colour of the continuum, unaffected by the emission lines. There are 
58 known stars with 14.0 $< V <$ 16.5 and $U-B$ $<$ $-$0.70 in the FBS area; 
39 (67\%) have been found by the FBS; we shall adopt this value as 
the completeness of this survey for QSOs brighter than $V$ = 16.5.

The PG survey does not cover the region at galactic latitudes lower than 
30\degr. The PG and FBS samples have about 2\,250 deg$^{2}$ in common. Out of 
the 1\,103 FBS objects, 618 are within the PG fields, 276 being in the 
PG sample (FBS 0854+385 is PG 0854+385, but the original FBS R.A. is 
affected by a printing error; FBS 0935+395 is not a PG object; 
PG 0752+770, 0836+619, 1047+694, 1335+369, 1551+719, 1600+369, 1606+627, 
1620+648 and 1722+353 are FBS objects, but their original PG positions 
are affected by errors reaching several arcminutes; the declination of 
FBS 1559+369 is affected by a printing error of 1\arcmin; it is G\,180$-$23 
[21] and PG 1600+369; the declination of FBS 1619+648 is also affected by a 
printing error of 1\arcmin. Accurate positions for these objects have 
been measured and are given in Table 1). 

\begin{table}
\begin{center}
\caption{\small Journal of observations and relevant data for the observed 
FBS objects. References are to the original lists of the FBS.} 
\vspace{0.5cm} 
\begin{tabular}{p{2cm}rcllrrl}
\tableline
\tableline
FBS Name &$b\:$  & Ref. & Mag  & Type & Date\verb+   +& Exp. time 
& Classification \\
 	 &       &      &      &      &          & (min)\verb+  + &          \\
\tableline
0019+348 &$-$27.4&  (2) & 15.0 & B2   & 27.10.97 & 10\verb+   + & CV         \\
0028+441 &$-$18.3& (11) & 14.5 & B2e: & 27.10.97 & 15\verb+   + & sdB        \\
0140+427 &$-$18.9& (11) & 16.5 & N1e  & 27.10.97 & 20\verb+   + & sdB        \\
0306+333 &$-$21.1&  (2) & 14.7 & B1   & 27.10.97 & 20\verb+   + & CV:        \\	
0632+663 &  23.3 &  (5) & 16.0 & N3e: & 28.10.97 & 15\verb+   + & sdB        \\	
0649+716 &  25.9 &  (6) & 17.1 & N1   & 28.10.97 & 20\verb+   + & featureless\\
0732+396 &  25.1 &  (9) & 16.  & B2   & 28.10.97 & 20\verb+   + & Sey1 
$z$ = 0.118 \\	
1112+668 &  47.9 &  (5) & 17.  & B2a: & 26.05.98 & 20\verb+   + & QSO 
$z$ = 0.544 \\
1150+334 &  76.0 &  (2) & 16.2 & N2e: & 26.05.98 & 20\verb+   + & QSO 
$z$ = 1.40 \\
1235+699 &  47.4 &  (6) & 17.9 & N1e  & 26.05.98 & 20\verb+   + & QSO 
$z$ = 0.522? \\
1255+447 &  72.6 & (11) & 16.5 & B1   & 25.05.98 & 20\verb+   + & QSO 
$z$ = 0.300 \\
1315+447 &  71.9 & (11) & 17.0 & N1   & 25.05.98 & 20\verb+   + & DZ: \\
1324+448 &  71.1 & (11) & 17.0 & B1   & 26.05.98 & 20\verb+   + & QSO 
$z$ = 0.331? \\
1401+865 &  30.7 &  (8) & 16.2 & N1e: & 25.05.98 & 20\verb+   + & DZ \\
1449+440 &  75.6 & (12) & 16.0 & N1   & 26.05.98 & 20\verb+   + & F0 \\
1452+762 &  39.0 &  (7) & 16.0 & N2e  & 25.05.98 & 20\verb+   + & sdB \\
1500+752 &  39.5 &  (7) & 16.9 & B2a  & 26.05.98 & 20\verb+   + & DA: \\
1523+363 &  56.3 &  (3) & 16.1 & N1   & 26.05.98 & 20\verb+   + & F0 \\
1557+448 &  48.9 & (12) & 16.5 & de:  & 25.05.98 & 20\verb+   + & \hii\ 
$z$ = 0.0417 \\
1607+439 &  47.1 & (12) & 16.0 & s1e: & 26.05.98 & 20\verb+   + & F0: \\
1715+406 &  34.5 &  (6) & 16.0 & sd3e & 25.05.98 & 20\verb+   + & sdF: \\
1716+394 &  34.2 & (10) & 17.0 & N1e: & 25.05.98 & 20\verb+   + & F0 \\
1755+663 &  30.2 &  (5) & 16.3 & N2   & 26.05.98 & 20\verb+   + & F0 \\
1810+374 &  23.4 &  (3) & 15.7 & B2   & 26.05.98 & 20\verb+   + & sdA \\
1819+348 &  20.9 &  (3) & 14.8 & B1e: & 15.06.98 & 20\verb+   + & sdA \\
1822+414 &  22.4 & (12) & 14.5 & B1   & 25.05.98 & 20\verb+   + & sdB-O \\
1822+352 &  20.5 &  (3) & 15.8 & B2   & 25.05.98 & 20\verb+   + & DA \\
1833+447 &  21.5 & (12) & 15.5 & B1a  & 25.05.98 & 20\verb+   + & F0 \\
2149+425 &$-$8.7 & (12) & 13.5 & B1   & 27.10.97 &  5\verb+   + & sdB \\	
2249+391 &$-$17.9& (10) & 16.5 & N1e: & 27.10.97 & 20\verb+   + & F5  \\
2315+443 &$-$15.2& (12) & 17.  & N2e: & 27.10.97 & 20\verb+   + & F5  \\
\tableline
\tableline
\end{tabular}
\end{center}
\end{table}
 
Forty-six PG objects have not been found in the FBS, but ten are Markarian
objects, i.e. belong to the first part of the FBS. So 88\% (276/312) of the PG
objects have been discovered. The 36 undiscovered objects have been examined
on the FBS plates; 24 have a weak UV excess (the PG survey finds a significant
fraction of stars with $U-B$ $\sim$ $-$0.4, while the FBS is relatively 
unsensitive for $U-B$ $>$ $-$0.7; the others are fainter than B$\sim$16 
and are near the magnitude limit of rather poor plates). From this, we 
conclude that the FBS survey is $\sim$ 90\% complete for $U-B$ $<$ $-$0.5. 
This is significantly larger than the 67\% success rate 
obtained from the $UBV$ stars; it is probably due to the fact that, 
in principle,
the PG survey contains only objects brighter than $B$ = 16.2.

There are 25 PG QSOs in the FBS area (listed in Table 4), 23 of them have 
been found; the two exceptions are PG 0953+414 and PG 1112+431; the first
is on the very edge of the FBS plate, while the second is weak on the original
plate and has been missed; its APS magnitude is also quite weak ($O$ = 17.03).
This confirms that the FBS is very efficient in discovering bright QSOs. 
However, at low galactic latitudes, there is only one FBS QSO, suggesting a
very low success rate which could be due in part to Galactic extinction and
reddening and in part to crowding on the objective prism plates.

\subsection{The AGN content of the FBS}
Thirty-four FBS objects are listed as QSOs in the eight edition of the
V\'eron-Cetty \& V\'eron catalogue [55]. Two more (FBS 1102+347 and 
1147+673) have been shown to be QSOs [25] and six have been identified in 
the present paper (Table 2). There are therefore 42 known QSOs in the FBS, 
41 being at high galactic latitude ($\vert b \vert$ $>$ 30\degr).

At high galactic latitudes, all FBS objects associated with a ROSAT RASS-BSC
source have been identified. Among them, there are 25 QSOs with known $z$, and
one without, altogether 26. As about 60\% of all PG QSOs are RASS 
sources, and assuming that this is true for the FBS QSOs, we should have a
total of about 43 QSOs in the FBS catalogue (including the one without 
known $z$). This suggests that the number of QSOs still to be found in the 
FBS catalogue is very small as 42 have already been found.

All 114 AGNs from the PG survey have been observed at 5\,GHz with the VLA [28];
thirty five (30\%) have been detected with a flux density larger than 3\,mJy.
The same fraction (12/40) of the known FBS QSOs have been detected in the NVSS
survey, suggesting that the number of QSOs in the as yet spectroscopically
unobserved FBS objects is small and probably cannot exceed about 10, as the
fraction of radio-detected QSOs would then drop below 25\% and be significantly
lower than the corresponding fraction for the PG survey.

\subsection{Completeness of the PG survey}
Goldschmidt et al. [22] have found a systematic difference of 0.28 mag 
between the PG magnitudes and their own measurements for 25 PG stars, the PG
magnitudes being too bright; they suggested that this difference was due to a
zero-point error in the PG magnitude scale. The mean differences between the PG
and photoelectric magnitudes for 105 stars is equal to 0.00 mag; this does not
confirm the existence of an systematic offset in the PG scale. The quoted error
for the PG photographic $B$ magnitude is $\sigma$ = 0.29 mag [23]. 
The comparison with photoelectric magnitudes gives 
$\sigma$ = 0.37 mag (Fig.~4b). 

\begin{figure}[t]
\begin{center}
\resizebox{16.5cm}{!}{\includegraphics{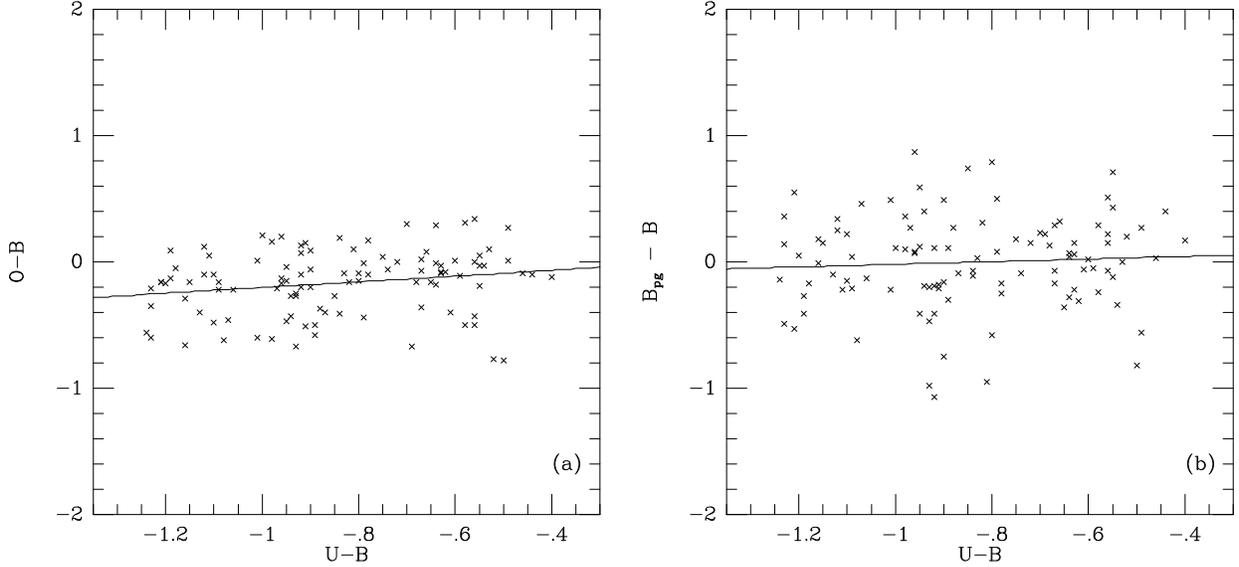}}
\caption{\small (a) Plot of the differences between 
the APS $O$ magnitudes and the photoelectric $B$ magnitudes for 105 PG objects. 
(b) Plot of the differences between the PG photographic and the photoelectric 
$B$ magnitudes for 
the same 105 objects. In the two figures, the straight lines represent 
the best fits through the data.}
\end{center}
\end{figure}

It has been suggested [58] that, on average, the PG magnitudes for QSOs are too
bright; as we do not observe such an effect for the stars, we suggest that this
is due to the QSO variability; QSOs are discovered preferentially when they are
bright; when measured at an epoch different than the survey epoch they are found
to be systematically weaker by a few tens of a magnitude [20].

\begin{table}
\begin{center}
\caption{\small PG and FBS QSOs in the FBS area (continues).} 
\vspace{0.5cm} 
\begin{tabular}{p{2cm}llcrccccl}
\tableline
\tableline
 Postion   &$\:\:\:z$& $\:\,B$ & $O$ & $M_{B}$ & &   &  & $b\:$  & Name \\
\tableline
 0732+396  & 0.118 & 16.:  &  14.70 & $-$24.6   & X & N & F  & 25.1 & \\
 0804+761  & 0.100 & 15.15 &  14.18 & $-$24.4   & X & Y & PF & 31.0 &  \\
 0838+770  & 0.131 & 16.30 &  17.55 & $-$22.0   & X & Y & PF & 32.7 &  \\
 0844+349  & 0.064 & 14.00 &  16.89 & $-$20.7   & X & Y & PF & 38.0 &  \\
 0931+437  & 0.456 & 16.41 &  16.47 & $-$25.8   & X & Y & PF & 47.4 & US 737 \\
 0935+416  & 1.966 & 16.30 &  16.07 & $-$29.6   & - & Y & PF & 48.3 &  \\
 0936+396  & 0.458 & 16.30 &  16.69 & $-$25.6   & - & Y & PF & 48.6 &  \\
 0947+396  & 0.206 & 16.40 &  16.39 & $-$24.1   & X & Y & PF & 50.7 &  \\
 0953+415  & 0.239 & 15.05 &  15.59 & $-$25.3   & X & Y & P  & 51.7 & \\
 0959+685  & 0.773 & 16.28 &  16.01 & $-$27.1   & x & Y & F  & 42.0 & \\
 1002+437  & 0.178 & 15.:  &  16.39 & $-$23.8   & X & Y & F  & 52.9 & \\
 1007+417  & 0.613 & 15.:  &  16.04 & $-$26.9   & X & Y & F  & 54.2 & 4C 41.21\\
 1048+342  & 0.167 & 15.81 &  15.94 & $-$24.2   & X & Y & PF & 63.4 & \\
 1049+617  & 0.421 & 16.66 &  16.62 & $-$25.3   & X & Y & F  & 50.4 & 4C 61.20\\
 1100+772  & 0.313 & 15.86 &  15.93 & $-$25.6   & X & Y & PF & 38.6 & 3C 249.1\\
 1102+347  & 0.51  & 16.2  &   --   & $-$26.2   & - & N & F  & 66.2 & CSO 314\\
 1112+668  & 0.544 & 17.0  &  16.53 & $-$26.1   & X & Y & F  & 47.9 & \\
 1112+431  & 0.302 & 16.20 &  17.03 & $-$24.4   & - & Y & P  & 64.9 & \\
 1114+445  & 0.144 & 16.05 &  15.11 & $-$24.6   & x & Y & PF & 64.5 & \\
 1115+407  & 0.154 & 16.02 &  14.57 & $-$25.3   & X & Y & PF & 66.7 & \\
 1121+422  & 0.234 & 16.02 &  15.84 & $-$24.6   & X & Y & PF & 66.9 & \\
 1137+661  & 0.650 & 16.50 &  16.25 & $-$26.6   & X & Y & F  & 49.7 & 3C 263.0\\
\tableline
\tableline
\end{tabular}
\vspace*{3mm}
\begin{tabular}{p{13cm}}
\footnotesize
X: in the ROSAT All Sky Survey Bright Source Catalogue; x: in the ROSAT WGACAT
Catalogue; R: in the NRAO VLA Sky Survey; Y: in the PG area; 
N: not in the PG area; O: APS $O$ magnitudes; P: in the PG catalogue; 
F: in the FBS catalogue.
\end{tabular}
\end{center}
\end{table}
\addtocounter{table}{-1}
\begin{table}
\begin{center}
\caption{\small PG and FBS QSOs in the FBS area (end).} 
\vspace{0.5cm} 
\begin{tabular}{p{2cm}llcrccccl}
\tableline
\tableline
 Position  &$\:\:\:z$& $\:\,B$ & $O$ & $M_{B}$ &  &   &  & $b\:$  & Name \\
\tableline
 1140+680  & 0.796 & 17.0  &  16.82 & $-$26.8   & X  & Y & F  & 48.1 & \\
 1147+673  & 1.02  & 16.7  &  16.69 & $-$27.2   & -- & Y & F  & 49.1 & \\
 1150+334  & 1.40  & 16.2  &  16.30 & $-$28.8   & R  & Y & F  & 76.0 & CSO 373\\
 1229+710  & 0.208 & 15.4  &  15.66 & $-$24.9   & X  & Y & F  & 46.3 & \\
 1235+699  & 0.522 & 17.9  &  17.96 & $-$24.5   & x  & Y & F  & 47.4 & \\
 1242+342  & 0.717 & 17.3  &  17.52 & $-$25.8   & -- & Y & F  & 83.1 & CSO 919\\
 1248+401  & 1.032 & 16.06 &  16.33 & $-$28.0   & X  & Y & PF & 77.3 & \\
 1255+447  & 0.300 & 16.5  &  16.48 & $-$24.9   & X  & Y & F  & 72.6 & \\
 1309+355  & 0.184 & 15.45 &  15.64 & $-$24.7   & X  & Y & PF & 80.7 & \\
 1322+659  & 0.168 & 15.86 &  15.71 & $-$24.2   & X  & Y & PF & 51.1 & \\
 1324+448  & 0.331 & 17.:  &  18.09 & $-$23.5   & X  & Y & F  & 71.1 & \\
 1329+412  & 1.937 & 16.30 &  16.78 & $-$29.1   & -- & Y & PF & 73.8 & \\
 1338+416  & 1.204 & 16.08 &  16.50 & $-$28.1   & -- & Y & PF & 72.5 & \\
 1351+640  & 0.088 & 15.42 &   --   & $-$22.3   & x  & Y & PF & 52.0 & \\
 1402+436  & 0.320 & 15.:  &   --   & $-$25.4   & -- & Y & F  & 68.0 & CSO 409 \\
 1411+442  & 0.089 & 14.99 &   --   & $-$22.6   & x  & Y & PF & 66.4 & \\
 1444+407  & 0.267 & 15.95 &   --   & $-$24.9   & X  & Y & PF & 62.7 & \\
 1512+370  & 0.370 & 15.97 &   --   & $-$25.8   & X  & Y & PF & 58.3 & B2 1512+37\\
 1526+659  & 0.345 & 17.0  &  16.90 & $-$24.8   & x  & Y & F  & 44.4 & \\
 1630+377  & 1.478 & 15.96 &  16.62 & $-$28.5   & x  & Y & PF & 42.9 & \\
 1634+706  & 1.337 & 14.90 &  15.27 & $-$29.7   & x  & Y & PF & 36.6 & \\
 1641+399  & 0.594 & 16.25 &  15.87 & $-$26.7   & X  & Y & F  & 40.9 & 3C 345.0\\ 
\tableline
\tableline
\end{tabular}
\vspace*{3mm}
\begin{tabular}{p{13.5cm}}
\footnotesize
X: in the ROSAT All Sky Survey Bright Source Catalogue; x: in the ROSAT WGACAT
Catalogue; R: in the NRAO VLA Sky Survey; Y: in the PG area; 
N: not in the PG area; O: APS $O$ magnitudes; P: in the PG catalogue; 
F: in the FBS catalogue.
\end{tabular}
\end{center}
\end{table}

The QSO counts are systematically affected by the photometric errors in $B$ as
these errors scatter many more objects toward brighter magnitudes than it does
toward fainter magnitudes. Assuming that the error distribution is Gaussian,
with dispersion $\sigma$, the correction to be applied 
to the observed counts is a factor 10 to the power of [$(b+1)\,\sigma^{2}/2$],  
where $b$ is the slope of the integrated number-magnitude relation: 
log\,N($B$) = $a + b \times B$ [18,53]. Assuming $b$ = 0.75, if 
$\sigma$ = 0.27 mag, the true QSO surface densities are smaller by 1.16 
than the observed ones; if $\sigma$ = 0.37, the correction is 1.32.

In principle the PG survey selected all objects with $U-B$ $<$ $-$0.46 
(and brighter than B$\sim$16.2); however, the $U-B$ colour was measured with 
a relatively large error (0.24 mag rms) which induced an incompleteness 
estimated at around 12\% [53]. Moreover, in the interval 0.6 $< z <$ 0.8, 
the strong \mgii\,$\lambda$2800 emission line is in the $B$ filter which 
results in a much redder $U-B$ colour than for 
neighbouring redshifts; as a result in this interval, the PG survey picked up
too few quasars and was estimated to be only 72\% complete [53].

The catalogue of mean $UBV$ data on stars [44] contains 283 stars in the
magnitude range 12.0 $< B <$ 16.5 and with $U-B$ $<$ $-$0.40 in the full 
10\,714 deg$^{2}$ area of the PG survey; 190 are included in the 
PG catalogue (there are 59 stars fainter than $B$ = 16.5 in the PG area, 
but only two are included in the PG catalogue). Twenty four stars 
photoelectrically observed because they were in the PG catalogue [29,37] 
have been ignored. The overall completeness of the PG survey is therefore 
64\% (166/259). 67\% (162/241) of the stars brighter than $B$ = 16.2 were 
found in the PG survey, while only 22\% (4/18) of those weaker than
this were detected. The completeness of the 
PG survey for stars brighter than $B$ = 16.2 
rises from about 55\% for $U-B$ $>$ $-$0.60 to 80\% for $U-B$ $<$ $-$1.0. 
For PG QSOs ($B <$ 16.2), the completeness should not be less than $\sim$70\%.
    
There are 19 known non-PG QSOs in the FBS catalogue, listed in Table 4; 17 are 
within the limits of the PG survey, but twelve have APS $O$ magnitudes 
weaker than
16.2 and may have been too weak for having been discovered by the PG survey.
FBS 1641+399 is 3C 345.0, an optically violently variable, with a $B$ magnitude
ranging from 14.7 to 17.7 [30]; according to the published light-curve, during
the epoch of the PG survey (Jan. 1973--May 1974) the object was always fainter
than $B$ = 16.4 and was most probably weaker than the plate limit. 
The published
magnitudes for FBS 1402+436 are inaccurate and in poor agreement 
($B$ = 15$\pm$0.75 [12], $B$ = 16 [49], $V$ = 16.5 [27]), suggesting that this 
object could have been weaker than the PG limiting magnitude. We are left 
with only three bright FBS QSOs missing from the PG catalogue 
implying an incompleteness of 15\% (assuming that the FBS is 70\% complete). 
Two of them are in the $z$ range (0.6--0.8) in which 
the $U-B$ excess is reduced because of the presence of the \mgii\ line in 
the $B$ filter which could explain their absence from the PG survey.

\begin{table}[h]
\begin{center}
\caption{\small Bright ($O <$ 16.2) QSOs at $\vert b \vert >$ 30\degr\ not 
in the FBS.} 
\vspace{0.5cm} 
\begin{tabular}{p{4cm}cllclccc}
\tableline
\tableline 
Name 		& Position & $\:\,z$ & $\:B$ &$O$&$\:\,M_{B}$ & &   &  $b$  \\
\tableline    
 HS 0806+6212   &  0806+62 & 0.173 & 16.5  &  16.12 & $-$24.0  & -- & Y & 33.0 \\
 KUV 08126+4154 &  0812+41 & 1.28  & 16.4  &  15.91 & $-$28.9  & -- & Y & 32.9 \\
 US 1329        &  0833+44 & 0.249 & 15.6  &  15.24 & $-$25.7  & X  & Y & 37.0 \\
 KUV 09468+3916 &  0946+39 & 0.360 & 16.1  &  15.99 & $-$25.8  & X  & Y & 50.6 \\
 RX J10265+6746 &  1022+68 & 1.178 & 15.0  &   --   & $-$29.6  & -- & Y & 43.9 \\
 KUV 11274+4133 &  1127+41 & 0.72  & 16.93 &  16.12 & $-$27.0  & -- & Y & 68.1 \\
 HS 1312+7837   &  1312+78 & 2.00  & 16.4  &  15.84 & $-$30.1  & -- & N & 38.7 \\
 CSO 1022       &  1351+36 & 0.284 & 16.   &   --   & $-$25.3  & -- & Y & 73.9 \\
 RX J14249+4214 &  1422+42 & 0.316 & 15.7  &   --   & $-$25.8  & X  & Y & 65.7 \\
 B3 1621+392    &  1621+39 & 1.97  & 16.7  &  15.86 & $-$30.0  & X  & Y & 44.7 \\
 RXS J16261+3359&  1624+34 & 0.204 & 16.5  &  15.82 & $-$24.7  & X  & Y & 43.8 \\
 RXS J17060+6857&  1706+69 & 0.449 & 16.3  &  16.04 & $-$26.2  & X  & N & 34.6 \\
 HS 1710+6753   &  1710+67 & 0.41  & 16.4  &  15.94 & $-$26.1  & -- & N & 34.5 \\
 B2 1721+34     &  1721+34 & 0.206 & 15.46 &   --   & $-$25.0  & X  & Y & 32.2 \\
\tableline
\tableline
\end{tabular}
\vspace*{3mm}
\begin{tabular}{p{13cm}}
\footnotesize
X: in the ROSAT All Sky Survey Bright Source Catalogue; Y: in the PG area; 
N: not in the PG area; O: APS $O$ magnitudes.
\end{tabular}
\end{center}
\end{table}

Wampler \& Ponz [58] suggested that the incompleteness of the PG survey could be
substantial. Goldschmidt et al. [22] found five new QSOs with $B <$ 16.17 in a
330 deg$^{2}$ area included in the PG area where Green et al. [23] found only
one; they got a surface density of 0.018 deg$^{-2}$, about three times 
larger than PG. We have obtained a spectrum of one of them (Q 1404$-$0455) 
which shows it to be a starburst galaxy at $z$ = 0.029. For two others, 
the $O$ magnitudes extracted from the APS database [48] are greater than 16.5; 
it is not clear if this is due to variability or to a difference in the 
magnitude scales. In these conditions, it seems hazardous to conclude to 
a gross incompleteness of the PG survey on the basis of these data. 
K\"ohler et al. [33] surveyed a 611 deg$^{2}$ area and concluded to an 
incompleteness of the BQS by a factor 2 to 3; they found eight QSOs brighter 
than $B$ = 16.16, or 0.013 deg$^{-2}$. La Franca \& Cristiani [36] 
have surveyed an area of 555 deg$^{2}$ in the magnitude range 
15 $< B <$ 18.75; they found that, for magnitudes brighter than $B$ = 16.4, 
the QSO surface density (0.013 deg$^{-2}$, derived from seven objects) is a 
factor 2.2 higher than the PG value. Savage et al. [51] found 16 QSOs 
brighter than $B$ = 16.16 in a 1\,500 deg$^{2}$ area or 0.011 deg$^{-2}$. 
These samples are quite small; the zero-point errors of their magnitude 
scales have not been determined. These results should, therefore, be 
considered as tentative.

\section{BUILDING A ``COMPLETE'' QSO SURVEY BASED ON APS $O$ MAGNITUDES}

Because of their variability, it is an impossible task to compare two QSO
surveys of the same region of the sky made at different epochs. However we
now have, for a large fraction of the sky, the possibility to extract from the
APS database, for any object, the $O$ magnitude as measured on the Palomar Sky
Survey plates [48] with an accuracy of about 0.2 mag [38]. By doing this for
all known QSOs found in the same area of the sky during a number of different
surveys, we may hope to get as near as possible from an ideal survey complete 
to a well defined limiting magnitude.

We have extracted the $O$ magnitudes of 105 PG UV-excess stars (excluding CVs).
We have compared these magnitudes with the photoelectric $B$ magnitudes [44] and
found a color equation: $O-B$ = $0.23\,(U-B)+0.02$ (Fig.~4a); 
the rms error on the $O$ magnitudes is 0.26 mag, slightly larger than 
the published value. For $U-B$ = $-$0.8, the mean value for QSOs, the $O$ 
magnitudes are systematically too bright by 0.16 mag.

We have extracted the APS $O$ magnitudes, when available, for all objects in
the QSO catalogue [55] brighter than $B$ = 17, with $M_{B} <$ $-$24.0 
and $z <$ 2.15, located in the 2\,400 deg$^{2}$ of the FBS at 
$\vert b \vert >$ 30\degr. Whenever this $O$ magnitude exists, we give 
it the preference. Table 4 contains 15 such QSOs with $O <$ 16.2 (and 11 
with $O <$ 16.0, corresponding to $B$ = 16.2) and three with
$B <$ 16.2. We have found ten additional QSOs with $O <$ 16.2 (seven with 
$O <$ 16.0) and four with $B <$ 16.2 (listed in Table 5).

Thus our ``complete'' sample contains between 18 and 25 QSOs brighter than
$B$ = 16.2, or 0.0075 to 0.010 deg$^{-2}$; this is 1.2 to 1.6 times larger 
than the PG surface density. If we correct these surface densities for the 
Eddington effect (1.16 for our survey and 1.32 for the PG survey), our 
surface densities are 1.4 to 1.8 times larger than the PG values.
 
It should be possible, when the APS database will be completed, to check the 
$O$ magnitudes of the seven objects for which they are not yet available.

\begin{acknowledgments}
ACKNOWLEDGMENTS~~A.M. Mickaelian is grateful to the CNRS for 
making possible his visit to Observatoire de Haute-Provence for 
carrying out this work. A.C. Gon\c{c}alves acknowledges support from 
the {\it Funda\c{c}\~ao para a Ci\^encia e a Tecnologia}, Portugal 
(PRAXIS XXI/BD/5117/95 PhD. grant).
\end{acknowledgments}
\vspace*{2cm}

\centerline{\bf REFERENCES}

 [1] ~Abrahamian H.V., Mickaelian A.M., Astrophysics 35, 363, 1991.

 [2] ~Abrahamian H.V., Mickaelian A.M., Astrophysics 36, 62, 1993.

 [3] ~Abrahamian H.V., Mickaelian A.M., Astrophysics 36, 306, 1993.

 [4] ~Abrahamian H.V., Mickaelian A.M., Astrophysics 37, 27, 1994.

 [5] ~Abrahamian H.V., Mickaelian A.M., Astrophysics 37, 117, 1994.

 [6] ~Abrahamian H.V., Mickaelian A.M., Astrophysics 37, 224, 1994.

 [7] ~Abrahamian H.V., Mickaelian A.M., Astrophysics 38, 108, 1995.

 [8] ~Abrahamian H.V., Mickaelian A.M., Astrophysics 39, 315, 1996.

 [9] ~Abrahamian H.V., Lipovetsky V.A., Stepanian J.A., Astrophysics 
   32, 14, 1990.

[10] Abrahamian H.V., Lipovetsky V.A., Mickaelian A.M., Stepanian J.A., 
   Astrophysics 33, 418, 1990.

[11] Abrahamian H.V., Lipovetsky V.A., Mickaelian A.M., Stepanian J.A., 
   Astrophysics 33, 493, 1990.

[12] Abrahamian H.V., Lipovetsky V.A., Mickaelian A.M., Stepanian J.A., 
   Astrophysics 34, 7, 1991.

[13] Allen S.W., Edge A.C., Fabian A.C. et al. , MNRAS 259, 67, 1992 

[14] Bade N., Engels D., Voges W., Astron. Astrophys. Suppl. Ser. 127, 145, 1998


[15] Beichman C.A., Neugebauer G., Habing H.J., Clegg P.E., Chester T.J. eds
   IRAS Catalogs and Atlases. 2. Point Source Catalog. Declination Range 
   90>Dec>30. Joint IRAS Science Working Group. NASA, Washington, DC: US 
   GPO, 1988.

[16] Condon J.J., Cotton W.D., Greisen E.W. et al., Astron. J. 115, 1693, 1998

[17] Downes R.A., Webbing R.F., Shara M.M., PASP 109, 345, 1997


[18] Eddington A.\,S., MNRAS 100, 354, 1940

[19] Eritsian M.\,A., Mickaelian A.\,M., Astrophysics 36, 126, 1993


[20] Francis P.\,J., 1996, Publ. astron. soc Australia 13, 212, 1996

[21] Giclas H.L., Burnham R., Thomas N.G., Lowell Obs. Bull. 129, 1965

[22] Goldschmidt P., Miller L., La Franca F., Cristiani S., MNRAS 256, 65P, 1992

[23] Green R.F., Schmidt M. , Liebert J., Astrophys. J. Suppl. 61,305, 1986.

[24] Gregory P.C., Scott W.K., Douglas K., Condon J.J., Astrophys. J. Suppl. 103,
   427, 1996

[25] Hagen H.-J., Engels D., Reimers D. ,A\&AS (in press), 1998

[26] Hayman P.G., Hazard C., Sanitt N., MNRAS 189, 853, 1979

[27] Hutchings J.B., Neff S.G., Astron. J. 104, 1, 1992

[28] Kellerman K.I., Sramek R., Schmidt M., Shaffer D.B., Green R., Astron. J.
   98, 1195, 1989

[29] Kidder K.M., Holdberg J.B., Mason P.A., Astron. J. 101, 579, 1991

[30] Kidger M.R., Astron. Astrophys. 226, 9, 1989

[31] Kilkenny D., Heber U., Drilling J.S., South African astron. obs. Circ. 12,
   1988

[32] King I.R., Raff M.I., PASP 89, 120, 1977

[33] K\"ohler T., Groote D. , Reimers D., Wisotzki L., Astron. Astrophys. 325, 
   502, 1997 

[34] Kondo M., Watanabe E., Yutani M., Noguchi T., Publ. astron. Soc. Japan 34,
   541, 1982

[35] Kondo M., Noguchi T., Maehara H., Ann. Tokyo Astron. Obs., 2nd Ser. 20, 130,
   1984

[36] La Franca F., Cristiani S., Astron. J. 113, 1517, 1997

[37] Landolt A.U., Astron. J. 104, 340, 1992

[38] Larsen J,A., Humphreys R.M., Astrophys. J. 436, L149, 1994

[39] Lemaitre G., Kohler D., Lacroix D., Meunier J.-P., Vin A., Astron. 
   Astrophys. 228, 546, 1989.

[40] Markarian B.E., Lipovetsky V.A., Stepanian J.A., Erastova L.K., Shapovalova
   A.I., Commun. Special Astrophys. Obs. 62, 5, 1989.

[41] Marsh M.C., Barstow M.A., Buckley D.A. et al., MNRAS 286, 369, 1997

[42] Massey P., Strobel K., Barnes J.V., Anderson E., Astrophys. J. 328, 315, 
   1988.

[43] McCook G.P., Sion E.M., Astrophys. J. Suppl. Ser. 65, 603, 1987


[44] Mermilliod J.-C., Mermilliod M., Catalogue of mean $UBV$ data on stars,
   Springer-Verlag,1994

[45] Mickaelian A.M., Discovery and Investigation of Blue Stellar Objects of the
   First Byurakan Survey, Ph.D. thesis, Byurakan, 1994.

[46] Mickaelian A.M., Eritsian M.A., Abramian G.V., Astrophysics 34, 186, 1991

[47] Noguchi T., Maehara H., Kondo M., Ann. Tokyo astron. Obs. 2nd ser. 18, 55,
   1980

[48] Pennington R.L., Humphreys R.M., Odewahn S.C., Zumach W., Thurmes P.M.,
   Publ. Astron. Soc. Pacific 105, 521, 1993

[49] Pesch P., Sanduleak N., Astrophys. J. Suppl. Ser. 70, 163, 1989

[50] Rengelink R.B., Tang Y., de Bruyn A.G. et al., Astron. Astrophys. Supl. Ser.
   124, 259, 1997

[51] Savage A., Cannon R.D., Stobie R.S. et al., Proceedings ASA 10, 265, 1993

[52] Sanduleak N., Pesch P., Astrophys. J. Suppl. Ser. 66, 387, 1988

[53] Schmidt M., Green R.F., Astrophys. J. 269, 352, 1983

[54] V\'eron-Cetty M.-P., V\'eron P., Astron. Astrophys. 115, 97, 1996.

[55] V\'eron-Cetty M.-P., V\'eron P., ESO Scientific Report N$^{\rm o}$ 18, 
1998.

[56] Voges W., Aschenbach B., Boller T. et al., IAU circ. 6420, 1996

[57] Wagner R.M., Sion E.M., Liebert J., Starrfield S.G., Astrophys. J. 328, 213,
   1988

[58] Wampler E.J.,Ponz D., Astrophys. J. 298, 448, 1985

[59] Wegner G., Boley F.I., Astron. J. 105, 660, 1993

[60] Wegner G., McMahan R.K., Astron. J. 90, 1511, 1985

[61] Wegner G., McMahan R.K., Astron. J. 91, 139, 1986

[62] Wegner G., McMahan R.K., Astron. J. 96, 1933, 1988

[63] Wegner G., Swanson S.R., Astron. J. 99, 330, 1990

[64] Wegner G., Swanson S.R., Astron. J. 100, 1274, 1990

[65] Wegner G., McMahan R.K., Boley, Astron. J. 94, 1271, 1987

[66] White N.E., Giommi P., Angelini L., IAU circ. 6100, 1994

\end{document}